\documentclass[prb,twocolumn,showpacs]{revtex4}
\usepackage[dvips]{epsfig}
\usepackage[english]{babel}
\usepackage{graphicx}
\usepackage{amsmath,float}
\usepackage{amsfonts}
\usepackage[normalem]{ulem}
\usepackage[usenames]{color}

\makeatletter
\renewcommand \thetable {\@Roman\c@table}
\makeatother

\begin{document}
\title{Chemical potential in disordered organic materials}
\author{A. Sharma$\footnote[1]{Author to whom correspondence should be addressed.
Electronic mail: a2.sharma@vu.nl}$ and M. Sheinman}

\affiliation{Department of Physics and Astronomy, VU University, Amsterdam, The Netherlands}

\date{\today}

\begin{abstract}
Charge carrier mobility in disordered organic materials is being actively studied, motivated by several applications such as organic light emitting diodes and organic field-effect transistors. It is known that the mobility in disordered organic materials depends on the chemical potential which in turn depends on the carrier concentration. However, the functional dependence of chemical potential on the carrier concentration is not known. In this study, we focus on the chemical potential in organic materials with Gaussian disorder. We identify three cases of non-degenerate, degenerate and saturated regimes. In each regime we calculate analytically the chemical potential as a function of the carrier concentration and the energetic disorder from the first principles.
\end{abstract}

\maketitle

Charge transport in disordered organic materials is exploited in a wide range of devices, including organic light emitting diodes (OLEDs),\cite{salaneck1999} organic field-effect transistors (OFETs),\cite{leeuw1998} photoreceptors,\cite{weiss} and photovoltaic cells.\cite{hummelen2001} In disordered organic materials, the carrier mobility is due to thermally assisted tunneling "hopping" between localized molecular states.\cite{Baessler1993, Coehoorn2005} It is know that the carrier mobility depends on the temperature, energetic disorder, and carrier concentration.\cite{pasveer2005, breda2007} For efficient device modeling it is useful to have a compact analytical expression for mobility. An attempt was made by Coehoorn \textit{et. al}\cite{Coehoorn2005} where they provided an analytical expression for mobility in organic materials with Gaussian energetic disorder. However, the authors noted that the expression for mobility is not suitable for practical numerical device modeling because no analytical expression for the chemical potential as a function of carrier concentration and energetic disorder is available. The main purpose of this paper is to show that an analytical expression for the chemical potential in organic materials with Gaussian energetic disorder can be obtained from the first principles, with no free parameters. We show that the derived expression for the chemical potential is fairly accurate and the error involved is well below the thermal energy.

Within the Gaussian disorder model, it is assumed that the density of states (DOS) is given by:
\begin{equation}
g(E)=\frac{N_t}{\sqrt{2\pi}\sigma}\exp\left(-\frac{E^2}{2\sigma^2}\right),
\end{equation}
where $\sigma$ is the standard deviation of the DOS and is a measure of the energetic disorder while $N_t$ is the total number of hopping
sites per unit volume. For a given charge carrier concentration, $p$, the chemical potential, $\mu$, is related to the density in the following way:
\begin{equation}
\int\limits _{-\infty}^{\infty}\frac{g(E)}{1+\exp\left(\frac{E-\mu}{k_{\rm B}T}\right)}dE = pN_t,
\label{TheEquation}
\end{equation}
where $k_{\rm B}$ is the Boltzmann's constant and $T$ is the temperature.

In general, for given $p$ and $\sigma$, Eq.~\eqref{TheEquation} is solved iteratively to obtain the chemical potential. However, in the limit of vanishing carrier concentration, an analytical expression for the chemical potential can be easily obtained because in the limit of small $p$ carriers can be considered essentially independent of each other. In this case Eq.~\eqref{TheEquation} can be solved for $\mu$ by replacing the Fermi-Dirac (FD) statistics with the Boltzmann statistics.~\cite{Coehoorn2005} In this limit, referred to as the Boltzmann approximation (BA), the density of occupied states (the product of DOS and the FD distribution function, DOOS) is to a very good approximation given by a Gaussian, centered at the energy value $E = -\sigma^2/(k_{\rm B}T)$ .\cite{Coehoorn2005} The corresponding expression for $\mu$ is given by:
\begin{equation}
\mu = - \frac{\sigma ^2}{2k_{\rm B}T} + k_{\rm B}T\ln p .
\end{equation}
The BA limit is applicable to OLEDs, where under typical operating conditions, the concentration is 10$^{-4}$-10$^{-5}$ carriers per hopping site.\cite{tanase2003} On the other hand in OFETs~\cite{tanase2003} application of high gate voltage can lead to a concentration of 0.01-0.1 carriers per hopping site. At these carrier concentrations, interaction between carriers becomes significant implying that Pauli's exclusion principle must be taken into account. In this high concentration regime, the Boltzmann approximation is no longer valid and Fermi-Dirac distribution must be used. We show that even in this regime, an analytical solution for $\mu$ can be obtained in a simple and intuitive manner. In the following text we identify three regimes referred to as the non-degenerate, degenerate, and saturated regime. In each regime we calculate analytically the chemical potential as function of the carrier concentration.

Our objective is to solve Eq.~\eqref{TheEquation} for the chemical potential, $\mu$, for given values of carrier concentration, $p$, energetic disorder $\sigma$, and temperature $T$. We evaluate the integral in Eq.~\eqref{TheEquation} using a saddle point approximation. We assume that the saddle point, $E_*$, of the integrand is known. Assuming this one gets two coupled algebraic equations for the chemical potential, $\mu$, and the saddle point of the integrand instead of one integral equation for the chemical potential.\cite{Sheinman2011How} The first algebraic equation is obtained by imposing the condition that $E_*$ is the saddle point of the integrand in Eq.~\eqref{TheEquation}. On doing so we get the following equation for the chemical potential
\begin{equation}
\mu=E_*+k_{\rm B}T \ln\left( -\frac{\sigma^2}{E_*k_{\rm B}T }-1\right).
\label{muSaddlePointA}
\end{equation}
Saddle point approximation implies that the integrand (equivalently the DOOS) is a Gaussian centered at $E_*$ with a standard deviation $ \sigma_* $. Namely, Eq. \eqref{TheEquation} is approximated by
\begin{equation}
\int\limits _{-\infty}^{\infty}h(E)dE = pN_t,
\label{TheEquationApp}
\end{equation}
where the expression for the DOOS is
\begin{equation}
h(E) = \frac{N_te^{-\frac{E_*^2}{2 \sigma^2}} \left(E_*k_{\rm B}T+\sigma^2\right)}{\sqrt{2 \pi } \sigma^3}\exp\left[-\frac{(E-E_*)^2}{2 \sigma_*^2}\right],
\label{SaddlePointApp}
\end{equation}
and
\begin{equation}
\sigma_*=\sqrt{\frac{k_{\rm B}T\sigma^4}{\sigma^2(k_{\rm B}T-E_*)-E_*^2 k_{\rm B}T }}.
\end{equation}

On evaluating Eq.~\eqref{TheEquationApp} we obtain the second algebraic equation:

\begin{equation}
\frac{e^{-\frac{E_*^2}{2\sigma^2}}\left(E_*k_{\rm B}T +\sigma^2\right)}{\sigma\sqrt{\sigma^2-E_*^2-\frac{E_*\sigma^2}{k_{\rm B}T }}}= p.
\label{SaddlePointA}
\end{equation}

\begin{figure}[tp]
\includegraphics[width=\columnwidth]{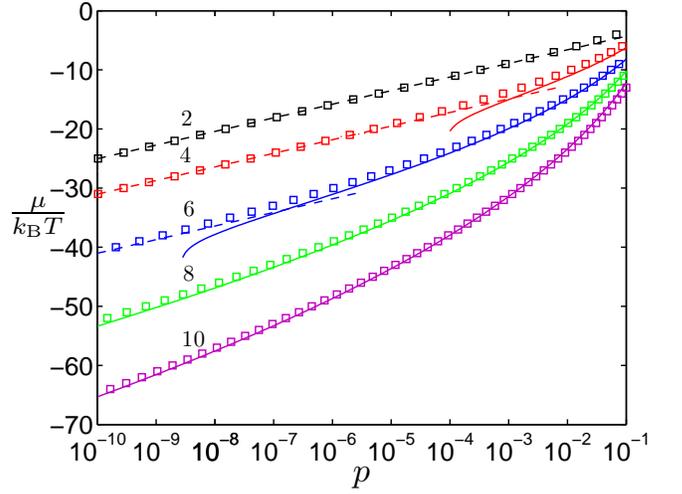}
\caption{Chemical potential $\mu$ vs. the normalized carrier concentration $p$ for different values of $\sigma/k_{\rm B}T$ (numbers on the plot). Full lines and dashed lines correspond to the non-degenerate, Eq. \eqref{muSelfAveA}, and degenerate, Eq. \eqref{muFreezeA}, regime, respectively. Symbols correspond to the numerically calculated chemical potential.
\label{Fig1}}
\end{figure}

\begin{figure}[b]
\includegraphics[width=\columnwidth]{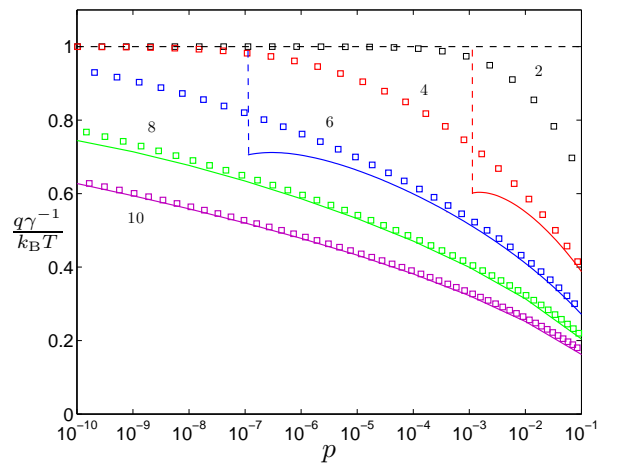}
\caption{Comparison of analytical result, Eq. \eqref{AnalyticGamma} (lines), and the numerical calculation (squares). Full lines correspond to the degenerate regime. The horizontal dotted line corresponds to the limit of the non-degenerate regime.
\label{Fig2}}
\end{figure}

\begin{figure}[tb]
\includegraphics[width=\columnwidth]{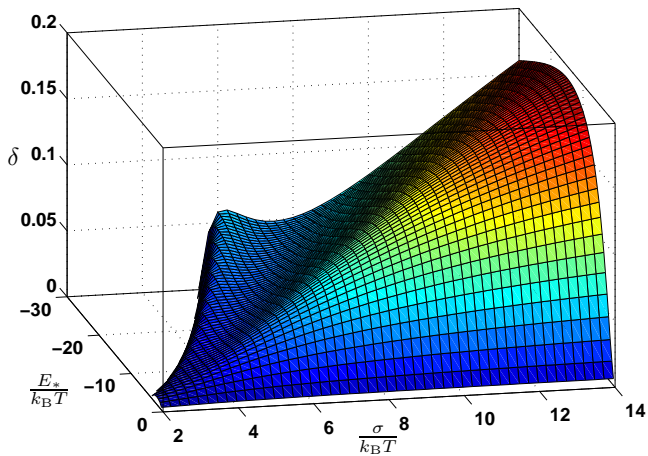}
\caption{Relative error $\delta$. For the range of $\sigma$ considered the error does not exceed $20\%$.
\label{Fig3}}
\end{figure}
The two coupled equations Eq.~\eqref{muSaddlePointA} and Eq.~\eqref{SaddlePointA} can be solved to obtain $\mu$ and $E_*$.
We now solve Eq.~\eqref{muSaddlePointA} and Eq.~\eqref{SaddlePointA} self-consistently in three limiting cases:

\paragraph{Non-degenerate regime:}
This regime corresponds to the case of vanishing carrier concentration. Eq.~\eqref{SaddlePointA} implies that in this regime the saddle point occurs at 
\begin{equation}
E_* \simeq -\sigma^2/k_{\rm B}T . 
\label{NonDegenerateSaddlePoint}
\end{equation}
The dependence of $E_*$ on the carrier concentration $p$ is obtained as
\begin{equation}
E_*=-\frac{\sigma^2}{k_{\rm B}T}\left(1- p e^{\sigma^2/2(k_{\rm B}T)^2}\right) .
\end{equation}
Self-consistency in this regime requires that carrier concentration is low enough such that the condition \eqref{NonDegenerateSaddlePoint} holds. This implies that in this regime $p \ll p_{1}$, where
\begin{equation}
p_1 = e^{-\frac{1}{2}\left(\frac{\sigma}{k_{\rm B}T}\right)^2}.
\label{p1}
\end{equation}
The chemical potential in this regime is well known and is given by
\begin{equation}
\mu=-\frac{\sigma^2}{2k_{\rm B}T} + k_{\rm B}T \ln p.
\label{muSelfAveA}
\end{equation}
The non-degenerate regime corresponds to the BA limit. In this regime an increase in carrier concentration leads to an increase of the maximum of DOOS with negligible shift in $E_*$.\cite{Coehoorn2005}

\paragraph{Degenerate regime:}
For $p \gg p_1$ the saddle point, $E_*$, shifts significantly from $ -\sigma^2/k_{\rm B}T  $ and the BA limit is no longer valid. In this regime the carrier concentration is such that saddle point satisfies $-\sigma^2/k_{\rm B}T \ll E_*  \ll -k_{\rm B}T $, so Eq. \eqref{SaddlePointA} implies
\begin{equation}
E_*=-\frac{\sigma}{\sqrt{2}} \sqrt{W\left[\frac{2}{ (\sigma/k_{\rm B}T)^2 p^4}\right]},
\end{equation}
where $W$ is the Lambert $W$-function. \cite{AS}
Self-consistency in this regime requires $p_1\ll p \ll p_2$, where
\begin{equation}
p_2=e^{-\frac{1}{2}\left(\frac{k_{\rm B}T }{\sigma}\right)^2}.
\end{equation}
The chemical potential in this regime is given by
\begin{widetext}
\begin{equation}
\mu=-\frac{\sigma}{\sqrt{2}} \sqrt{W\left[\frac{2}{ (\sigma/k_{\rm B}T)^2}\left(\frac{1}{p}\right)^4\right]}+k_{\rm B}T\ln \left\lbrace   \frac{\sqrt{2}(\sigma/k_{\rm B}T)}{\sqrt{W\left[\frac{2}{ (\sigma/k_{\rm B}T)^2}\left(\frac{1}{p}\right)^4\right]}}-1\right\rbrace.
\label{muFreezeA}
\end{equation}
\end{widetext}
This analytical expression for the chemical potential in the degenerate regime is the main result of this paper. Since our focus is on the systems with $ \sigma>k_{\rm B}T $, the second crossover value, $ p_2 $ is of the order of one and degenerate regime is valid until almost the full saturation of all the available states. For the sake of completeness we discuss below this saturated regime.

\paragraph{Saturated regime:}
In this regime the carrier concentration, $ p $, approaches the limiting value of $p_2 \simeq 1$ where almost all the states are occupied and the DOOS is the same as the density of states. In this limit, the saddle point satisfies $ 0 > E_*  \gg -k_{\rm B}T $, so Eq. \eqref{SaddlePointA} implies
\begin{equation}
E_*=-\sigma \sqrt{2\ln \frac{1}{p}}
\end{equation}
Self-consistency in this regime requires $p > p_2$.
The chemical potential in this regime is given by
\begin{equation}
\mu=-\sigma \sqrt{2\ln\frac{1}{p}}+k_{\rm B}T\ln\left(\frac{\sigma}{k_{\rm B}T\sqrt{2 \ln\frac{1}{p}}}-1\right)
\label{muSaturatedA}
\end{equation}
This regime is irrelevant for organic devices because even in OFETs, where carrier concentration can be tuned by applying gate voltage, the concentration is typically much smaller than $1$.\cite{tanase2003}

In Fig.~1, we plot the chemical potential as obtained above for different values of $\sigma$. As predicted by Eq. \eqref{p1}, with increasing $\sigma$, the crossover from the non-degenerate to the degenerate regime occurs at lower concentrations. The crossover from the degenerate to the saturated regime is not shown in the figure. The saturated regime, irrelevant for organic devices, is a narrow regime with divergence of chemical potential at $p=1$. It is clear from the figure that the analytically obtained chemical potential and the crossover density, Eq. \eqref{p1}, are in good agreement with the numerical calculations.

In addition to the mobility of carriers, to model charge transport one needs to calculate the diffusion coefficient. Assuming equilibrium conditions, the diffusion coefficient is determined from the the generalized Einstein relation\cite{ashcroft} which is given as
\begin{equation}
\gamma = \frac{p}{q}\frac{\partial \mu}{\partial p},
\label{GER}
\end{equation}
where $\gamma$ is ratio of diffusion coefficient to the mobility of carriers and $q$ is the elementary charge. Roichamn and Tessler\cite{tessler2002} showed that in a Gaussian DOS, $\gamma$ deviates significantly from the low density limit value of $k_{\rm B}T/q$. They obtained $\gamma$ as an implicit function of the chemical potential, $\mu$. Our expression for the chemical potential allows us to obtain an analytical expression for the generalized Einstein relation in each of the regimes as a function of $ p $, $ \sigma $ and $ T $:
\begin{widetext}
\begin{equation}
\gamma=\frac{k_{\mathrm{B}}T}{q}\left\{ \begin{array}{cc}
1 & p\ll p_{1}\\
\frac{\frac{\sqrt{2}\sigma}{k_{\mathrm{B}}T}\left\{ 2-W\left[\frac{2}{p^{4}\left(\sigma/k_{\mathrm{B}}T\right)^{2}}\right]+\frac{\sqrt{2}\sigma}{k_{\mathrm{B}}T}\sqrt{W\left[\frac{2}{p^{4}\left(\sigma/k_{\mathrm{B}}T\right)^{2}}\right]}\right\} }{\left\{ \frac{\sqrt{2}\sigma}{k_{\mathrm{B}}T}-\sqrt{W\left[\frac{2}{p^{4}\left(\sigma/k_{\mathrm{B}}T\right)^{2}}\right]}\right\} \left\{ 1+W\left[\frac{2}{p^{4}\left(\sigma/k_{\mathrm{B}}T\right)^{2}}\right]\right\} } & p_{2}\gg p\gg p_{1}\\
\frac{1+\frac{\sqrt{2}\sigma}{k_{\mathrm{B}}T}\sqrt{\ln\frac{1}{p}}-2\ln\frac{1}{p}}{2\ln\frac{1}{p}-2\sqrt{2}\text{ }\frac{k_{\mathrm{B}}T}{\sigma}\ln^{3/2}\frac{1}{p}} & p> p_{2}
\end{array}\right.
\label{AnalyticGamma}
\end{equation}
\end{widetext}
In Fig.~2 we compare our analytical expression for $ \gamma $, Eq. \eqref{AnalyticGamma}, with the numerical results (notice that the inverse of $\gamma$ is shown). As expected, a fairly good agreement is obtained far from the crossover concentration, $ p_1 $.

The results above indicate that the analytical expressions for the chemical potential are fairly accurate. Nevertheless, it is useful to estimate analytically the error in the derived formulas. The main source of the error is the saddle point approximation, Eq. \eqref{SaddlePointApp}, in evaluating the integral in Eq.~\eqref{TheEquation}. Under the saddle point approximation, the integrand (or equivalently the DOOS) of the integral in Eq.~\eqref{TheEquation} is a Gaussian centered at $E_*$. The integrand is well approximated by a Gaussian only in the neighborhood of $E_*$ (within few standard deviations, $\sigma_*$). Away from $E_*$, the deviation of the integrand from the Gaussian gives rise to the error in evaluation of Eq.~\eqref{TheEquation}. The relative error in saddle point approximation of the integral can be written as
\begin{equation}
\delta=\left|1-\frac{\int\limits _{-\infty}^{\infty}h\left( E\right) dE
}{\int\limits _{-\infty}^{\infty}\frac{g(E)}{1+\exp\left(\frac{E-\mu}{k_{\rm B}T}\right)}}\right|.
\end{equation}
For energies well below the saddle point, $E \ll E_*-3\sigma_*$, the FD distribution can be approximated as unity whereas well above the saddle point, $E \gg E_*+3\sigma_*$, it can be approximated by the Boltzmann distribution. Therefore the error can be estimated as
\begin{widetext}
 \begin{equation}
\delta\simeq\left|1-\frac{\int\limits _{-\infty}^{\infty}h\left( E\right) dE
}{\int\limits _{-\infty}^{E_*-3\sigma_*}g\left( E\right) dE+\int\limits _{E_*-3\sigma_*}^{E_*+3\sigma_*}h\left( E\right)dE+\int\limits _{E_*+3\sigma_*}^{\infty}g\left( E\right)e^{-\frac{E-\mu}{k_{\rm B}T}}dE}\right|.
\end{equation}
\end{widetext}

The integrals appearing in the above expression can be simplified using error function. For given values of carrier concentration, $p$, and energetic disorder, $\sigma$, the other parameters $E_*$, $\sigma_*$, and $\mu$ can be calculated as shown above. In Fig.~3 we show that for relevant range of $\sigma$ and $p$ values, the value of $\delta$ does not exceed 20\%. This implies that the error in the chemical potential is well below than the thermal energy, $k_{\rm B}T$.

Although we derived the expressions for the chemical potential only in the respective regimes, the algebraic equations Eq.~\eqref{muSaddlePointA} and Eq.~\eqref{SaddlePointA} can be easily solved numerically to any desired degree of accuracy for any given values of $p$ and $\sigma$.

To conclude, in this paper we derive from the first principles an analytical expression for the chemical potential in organic materials with Gaussian disorder. In the most relevant, degenerate regime our result, Eq. \eqref{muFreezeA}, is new and, as we demonstrate, is fairly accurate. We show that over the relevant range of carrier concentration and energetic disorder, the error in calculation is well below the thermal energy. In addition, the existing iterative numerical techniques to calculate chemical potential can use our derived expressions as fairly accurate starting point. This can lead to much more efficient algorithms to calculate chemical potential and therefore the mobility in organic materials with Gaussian disorder.

We thank M. Depken and P. A. Bobbert for helpful discussions.

\bibliography{references}

\begin{thebibliography}{13}
\expandafter\ifx\csname natexlab\endcsname\relax\def\natexlab#1{#1}\fi
\expandafter\ifx\csname bibnamefont\endcsname\relax
  \def\bibnamefont#1{#1}\fi
\expandafter\ifx\csname bibfnamefont\endcsname\relax
  \def\bibfnamefont#1{#1}\fi
\expandafter\ifx\csname citenamefont\endcsname\relax
  \def\citenamefont#1{#1}\fi
\expandafter\ifx\csname url\endcsname\relax
  \def\url#1{\texttt{#1}}\fi
\expandafter\ifx\csname urlprefix\endcsname\relax\def\urlprefix{URL }\fi
\providecommand{\bibinfo}[2]{#2}
\providecommand{\eprint}[2][]{\url{#2}}

\bibitem[{\citenamefont{Friend et~al.}(1999)\citenamefont{Friend, Gymer,
  Holmes, Burroughes, Marks, Taliani, Bradley, Dos~Santos, Br\'{e}das,
  L\"{o}gdlund et~al.}}]{salaneck1999}
\bibinfo{author}{\bibfnamefont{R.~H.} \bibnamefont{Friend}},
  \bibinfo{author}{\bibfnamefont{R.~W.} \bibnamefont{Gymer}},
  \bibinfo{author}{\bibfnamefont{A.~B.} \bibnamefont{Holmes}},
  \bibinfo{author}{\bibfnamefont{J.~H.} \bibnamefont{Burroughes}},
  \bibinfo{author}{\bibfnamefont{R.~N.} \bibnamefont{Marks}},
  \bibinfo{author}{\bibfnamefont{C.}~\bibnamefont{Taliani}},
  \bibinfo{author}{\bibfnamefont{D.~D.~C.} \bibnamefont{Bradley}},
  \bibinfo{author}{\bibfnamefont{D.~A.} \bibnamefont{Dos~Santos}},
  \bibinfo{author}{\bibfnamefont{J.~L.} \bibnamefont{Br\'{e}das}},
  \bibinfo{author}{\bibfnamefont{M.}~\bibnamefont{L\"{o}gdlund}},
  \bibnamefont{et~al.}, \bibinfo{journal}{Nature(London)}
  \textbf{\bibinfo{volume}{397}}, \bibinfo{pages}{121} (\bibinfo{year}{1999}).

\bibitem[{\citenamefont{Drury et~al.}(1998)\citenamefont{Drury, Mutsaers, Hart,
  Matters, and de~Leeuw}}]{leeuw1998}
\bibinfo{author}{\bibfnamefont{C.~J.} \bibnamefont{Drury}},
  \bibinfo{author}{\bibfnamefont{C.~M.~J.} \bibnamefont{Mutsaers}},
  \bibinfo{author}{\bibfnamefont{C.~M.} \bibnamefont{Hart}},
  \bibinfo{author}{\bibfnamefont{M.}~\bibnamefont{Matters}}, \bibnamefont{and}
  \bibinfo{author}{\bibfnamefont{D.~M.} \bibnamefont{de~Leeuw}},
  \bibinfo{journal}{Appl. Phys. Lett.} \textbf{\bibinfo{volume}{73}},
  \bibinfo{pages}{108} (\bibinfo{year}{1998}).

\bibitem[{\citenamefont{Borsenberger and Weiss}(1998)}]{weiss}
\bibinfo{author}{\bibfnamefont{M.}~\bibnamefont{Borsenberger}}
  \bibnamefont{and} \bibinfo{author}{\bibfnamefont{D.~S.} \bibnamefont{Weiss}},
  \emph{\bibinfo{title}{Organic photoreceptors for Xeroxgraphy}}
  (\bibinfo{publisher}{Marcel Dekker, New York}, \bibinfo{year}{1998}).

\bibitem[{\citenamefont{Brabec et~al.}(2001)\citenamefont{Brabec, Sariciftci,
  and Hummelen}}]{hummelen2001}
\bibinfo{author}{\bibfnamefont{C.~J.} \bibnamefont{Brabec}},
  \bibinfo{author}{\bibfnamefont{N.~S.} \bibnamefont{Sariciftci}},
  \bibnamefont{and} \bibinfo{author}{\bibfnamefont{J.~C.}
  \bibnamefont{Hummelen}}, \bibinfo{journal}{Adv. Funct. Mater.}
  \textbf{\bibinfo{volume}{11}}, \bibinfo{pages}{15} (\bibinfo{year}{2001}).

\bibitem[{\citenamefont{B\"assler}(1993)}]{Baessler1993}
\bibinfo{author}{\bibfnamefont{H.}~\bibnamefont{B\"assler}},
  \bibinfo{journal}{Phys. Stat. Sol. B} \textbf{\bibinfo{volume}{175}},
  \bibinfo{pages}{15} (\bibinfo{year}{1993}).

\bibitem[{\citenamefont{Coehoorn et~al.}(2005)\citenamefont{Coehoorn, Pasveer,
  Bobbert, and Michels}}]{Coehoorn2005}
\bibinfo{author}{\bibfnamefont{R.}~\bibnamefont{Coehoorn}},
  \bibinfo{author}{\bibfnamefont{W.~F.} \bibnamefont{Pasveer}},
  \bibinfo{author}{\bibfnamefont{P.~A.} \bibnamefont{Bobbert}},
  \bibnamefont{and} \bibinfo{author}{\bibfnamefont{M.~A.~J.}
  \bibnamefont{Michels}}, \bibinfo{journal}{Phys. Rev. B}
  \textbf{\bibinfo{volume}{72}}, \bibinfo{pages}{155206}
  (\bibinfo{year}{2005}).

\bibitem[{\citenamefont{Pasveer et~al.}(2005)\citenamefont{Pasveer, Cottaar,
  Tanase, Coehoorn, Bobbert, Blom, de~Leeuw, and Michels}}]{pasveer2005}
\bibinfo{author}{\bibfnamefont{W.~F.} \bibnamefont{Pasveer}},
  \bibinfo{author}{\bibfnamefont{J.}~\bibnamefont{Cottaar}},
  \bibinfo{author}{\bibfnamefont{C.}~\bibnamefont{Tanase}},
  \bibinfo{author}{\bibfnamefont{R.}~\bibnamefont{Coehoorn}},
  \bibinfo{author}{\bibfnamefont{P.~A.} \bibnamefont{Bobbert}},
  \bibinfo{author}{\bibfnamefont{P.~M.} \bibnamefont{Blom}},
  \bibinfo{author}{\bibfnamefont{D.~M.} \bibnamefont{de~Leeuw}},
  \bibnamefont{and} \bibinfo{author}{\bibfnamefont{M.~A.~J.}
  \bibnamefont{Michels}}, \bibinfo{journal}{Phys. Rev. Lett.}
  \textbf{\bibinfo{volume}{94}}, \bibinfo{pages}{206601}
  (\bibinfo{year}{2005}).

\bibitem[{\citenamefont{Coropceanu et~al.}(2007)\citenamefont{Coropceanu,
  Cornil, da~Silva~Filho, Olivier, Silbey, and Brédas}}]{breda2007}
\bibinfo{author}{\bibfnamefont{V.}~\bibnamefont{Coropceanu}},
  \bibinfo{author}{\bibfnamefont{J.}~\bibnamefont{Cornil}},
  \bibinfo{author}{\bibfnamefont{D.~A.} \bibnamefont{da~Silva~Filho}},
  \bibinfo{author}{\bibfnamefont{Y.}~\bibnamefont{Olivier}},
  \bibinfo{author}{\bibfnamefont{R.}~\bibnamefont{Silbey}}, \bibnamefont{and}
  \bibinfo{author}{\bibfnamefont{J.-L.} \bibnamefont{Brédas}},
  \bibinfo{journal}{Chem. Rev.} \textbf{\bibinfo{volume}{107}},
  \bibinfo{pages}{926} (\bibinfo{year}{2007}).

\bibitem[{\citenamefont{Tanase et~al.}(2003)\citenamefont{Tanase, Meijer, Blom,
  and de~Leeuw}}]{tanase2003}
\bibinfo{author}{\bibfnamefont{C.}~\bibnamefont{Tanase}},
  \bibinfo{author}{\bibfnamefont{E.~J.} \bibnamefont{Meijer}},
  \bibinfo{author}{\bibfnamefont{P.~W.~M.} \bibnamefont{Blom}},
  \bibnamefont{and} \bibinfo{author}{\bibfnamefont{D.~M.}
  \bibnamefont{de~Leeuw}}, \bibinfo{journal}{Phys. Rev. Lett.}
  \textbf{\bibinfo{volume}{91}}, \bibinfo{pages}{216601}
  (\bibinfo{year}{2003}).

\bibitem[{\citenamefont{{Sheinman} and {Kafri}}(2011)}]{Sheinman2011How}
\bibinfo{author}{\bibfnamefont{M.}~\bibnamefont{{Sheinman}}} \bibnamefont{and}
  \bibinfo{author}{\bibfnamefont{Y.}~\bibnamefont{{Kafri}}},
  \bibinfo{journal}{ArXiv e-prints}  (\bibinfo{year}{2011}),
  \eprint{1110.5997}.

\bibitem[{\citenamefont{Abramowitz and Stegun}(1964)}]{AS}
\bibinfo{author}{\bibfnamefont{M.}~\bibnamefont{Abramowitz}} \bibnamefont{and}
  \bibinfo{author}{\bibfnamefont{I.~A.} \bibnamefont{Stegun}},
  \emph{\bibinfo{title}{Handbook of Mathematical Functions with Formulas,
  Graphs, and Mathematical Tables}} (\bibinfo{publisher}{Dover publications,
  New York}, \bibinfo{year}{1964}).

\bibitem[{\citenamefont{Ashcroft and Mermin}(1988)}]{ashcroft}
\bibinfo{author}{\bibfnamefont{N.~W.} \bibnamefont{Ashcroft}} \bibnamefont{and}
  \bibinfo{author}{\bibfnamefont{N.~D.} \bibnamefont{Mermin}},
  \emph{\bibinfo{title}{Solid State Physics}} (\bibinfo{publisher}{Holt,
  Rinehart and Winston, New York}, \bibinfo{year}{1988}).

\bibitem[{\citenamefont{Roichman and Tessler}(2002)}]{tessler2002}
\bibinfo{author}{\bibfnamefont{Y.}~\bibnamefont{Roichman}} \bibnamefont{and}
  \bibinfo{author}{\bibfnamefont{A.}~\bibnamefont{Tessler}},
  \bibinfo{journal}{Appl. Phys. Lett.} \textbf{\bibinfo{volume}{80}},
  \bibinfo{pages}{1948} (\bibinfo{year}{2002}).

\end{thebibliography}
\end{document}